\documentclass{llncs}

\usepackage{makeidx}  

\usepackage{graphicx}
\usepackage{amsmath}
\usepackage{amsfonts}
\usepackage{amssymb}

\def\trait{\leavevmode\hskip2pt\raise2pt\hbox{\vrule%
        height 0.4pt depth 0pt width 4mm}\hskip2pt}
\def\M{{\rm I \! \kern-0.5pt M \kern0.5pt}(X,D)} 
\def\N{{\rm I \! \kern-0.5pt N \kern0.5pt}} 
\def\MM{{\rm I \! \kern-0.5pt M \kern0.5pt}}
\def\Mp{{\rm I \! \kern-0.5pt M^+ \kern0.5pt}(X,D)}
\def\Box{\setbox2=\hbox{$\sqcap$}%
\typeout{--- \the\wd2}\sqcap\!\!\!\!\hskip1pt\raise-0pt\hbox to6pt{\hrulefill}}

\begin{document}

%
\pagestyle{headings}  

\title{Methods for Partitioning Data to Improve Parallel Execution
Time for Sorting on Heterogeneous Clusters\thanks{Work supported in part by France Agence Nationale de la Recherche under grants ANR-05-SSIA-0005-01 and ANR-05-SSIA-0005-05, programme ARA s\'ecurit\'e}}

\author{Christophe C\'erin\inst{1}, Jean-Christophe Dubacq\inst{1}
\and Jean-Louis Roch\inst{2}}

%
\institute{Universit\'e de Paris Nord, LIPN, CNRS UMR 7030,\\ 99 avenue 
J.B Cl\'ement, 93430 Villetaneuse - France\\
\email{\{cerin,jcdubacq\}@lipn.univ-paris13.fr}
\and
ID-IMAG, CNRS - INRIA - INPG - UJF, Projet MOAIS\\
51 Av. J. Kuntzmann, 38330 Montbonnot-Saint-Martin - France\\
\email{Jean-Louis.Roch@imag.fr}}

\maketitle

\begin{abstract} 
  The aim of the paper is to introduce general techniques in order to
  optimize the parallel execution time of sorting on a distributed
  architectures with processors of various speeds. Such an application
  requires a partitioning step. For uniformly related processors
  (processors speeds are related by a constant factor), we develop a
  constant time technique for mastering processor load and execution
  time in an heterogeneous environment and also a technique to deal
  with unknown cost functions. For non uniformly related processors,
  we use a technique based on dynamic programming. Most of the time,
  the solutions are in ${\cal O}(p)$ ($p$ is the number of
  processors), independent of the problem size $n$. Consequently, there is
  a small overhead regarding the problem we deal with but it is inherently
  limited by the knowing of time complexity of the portion of code
  following the partitioning. 
  
\end{abstract}


The advent of parallel processing, in particular in the context of
{\it cluster computing} is of particular interest with the available
technology. A special class of {\it non homogeneous clusters} is under
concern in the paper. We mean clusters whose global performances are
correlated by a multiplicative factor.  We depict a cluster by the
mean of a vector set by the relative speeds of each processor.

In this paper we develop general techniques in order to control the
execution time and the load balancing of each node for applications
running in such environment. What is important over the application we
consider here, is the meta-partitioning schema which is the key of
success. All the approaches we develop can be considered as static
methods: we predetermine the size of data that we have to exchange
between processors in order to guarantee that all the processors end
at the same time before we start the execution. So, this work can be
considered in the domain of placement of tasks in an heterogeneous
environment. 

Many works have been done in data partitioning on heterogeneous
platforms, among them Lastovetsky's and Reddy's work \cite{ipdps04}
that introduces a scheme for data partitioning when memory hierarchies
from one CPU to another are different. There, the heterogeneity notion
is related to the heterogeneity of the memory structure. Under the
model, the speed of each processor is represented by a function of the
size of the problem.  The authors solve the problem of partitioning
$n$ elements over $p$ heterogeneous processors in ${\cal O}(p^2\times
\log_2 n)$ time complexity.

Drozdowski and Lawenda in \cite{europar2005} propose two algorithms
that gear the load chunk sizes to different communication and
computation speeds of applications under the principle of divisible
loads (computations which can be divided into parts of arbitrary
sizes; for instance painting with black pixels a whole image). The
problem is formalized as a linear problem solved either by branch and
bound technique or a genetic algorithm. Despite the fact that the
architecture is large enough (authors consider heterogeneous CPU and
heterogeneous links), we can not apply it here because our problem
cannot be expressed under the framework of 'divisible loads': in our
case, we need to merge sorted chunks after the partitioning step and
the cost is not a linear one\ldots thus our new technique.


The organization of our paper is the following. In section~\ref{sec:ta} we
introduce the problem of sorting in order to characterize the
difficulties of partitioning data in an heterogeneous environment. The
section motivates the work. In section~\ref{sec:previous}  we recall our previous
techniques and results. Section~\ref{sec:analytic} is devoted to a new constant time
solution and deals also with unknown cost functions. In section~\ref{sec:dynamic} we
introduce a dynamic programming approach and we recall a technique
that do not assume a model of processors related by constant integers
but in this case the processor speed may be ``unrelated''. Section~\ref{sec:experiments}
is about experiments and section~\ref{sec:conclusion} concludes the paper.

\section{Target applications and implementation on heterogeneous clusters}\label{sec:ta}

Assume that you have a set of $p$ processors with different speeds,
interconnected by a crossbar. Initially, the data is distributed
across the $p$ processors and according to the speeds: the slowest
processor has less data than the quickest. This assumption describes
the initial condition of the problem.  In this section we detail our
sorting application for which performance are directly related to this
initial partitioning.

\subsection{Parallel Sort.}\label{mysection}
Efficient parallel sorting on clusters (see
\cite{spaa94*46,reif87,STOC::ReifV1983,ShiHanmaoa1992a,LiXandLuPa1993a,HelmanDavi1996b}
for the homogeneous case and \cite{CG00,CG00a,CG00b,CG02,CERIN2002}
for the heterogeneous case) can be implemented in the following ways:
\begin{enumerate}

\item Each processor sorts locally its portion and picks up
representative values in the sorted list. It sends the representative
values to a dedicated node.

\item This node sorts what it receives from the processors and it keeps $p-1$ 
pivots; it distributes the pivots to all the processors.

\item Each processor partitions its sorted input according to the pivots
and it sends $p-1$ portions to the others.

\item Each processor merges what it received from the others.

\end{enumerate}
Note that the sorting in step 1 can be bypassed but in this case the
last step is a sort not a merge. Moreover note that there is only one
communication step: the representative values can be selected by
sampling few candidates at a cost much lower than the exchange of
values. In other words, when a value moves, it goes to the final
destination node in one step.

\section{Previous results and parallel execution time}\label{sec:previous}


Consider the simple problem of local sorting, such as presented in
\cite{CG00a} (and our previous comments). The sizes $n_i$ of data
chunks on each node is assumed to be proportional to the speed of
processors.

Let us now examine the impact on the parallel execution time of
sorting of the initial distribution or, more precisely, the impact of
the redistribution of data.  We determine the impact in terms of the
way of restructuring the code of the meta partitioning scheme that we
have introduced above. In the previous section, when we had $N$ data
to sort on $p$ processors depicted by their respective speeds
$k1,\cdots,k_p$, we had needed to distribute to processor $p_i$ an
amount $n_i$ of data such that:
\begin{equation}
n_1 / k_1 = n_2 / k_2 = .....= n_p / k_p
\end{equation}
and
\begin{equation}
n_1 + n_2 + .... + n_p = N
\end{equation}
The solution is:
$$
\forall i,n_i  = N\times k_i / (k_1+k_2+ ...+k_p)
$$

Now, since the sequential sorts are executed on $n_i$ data at a cost
proportional $n_i\ln n_i$ time cost (approximatively since there is a
constant in front of this term), there is no reason that the nodes
terminate at the same time since $n_1/k_1\ln n_1 \neq n_2/k_2\ln n_2
\neq\cdots \neq n_p/k_p\ln n_p$ in this case. The main idea that we
have developed in \cite{sbac04} is to send to each processor an amount
of data to be treated by the sequential sorts proportional to $n_i\ln
n_i$.  The goal is to minimize the global computation time $T = \min
(\max_{i=1, \ldots, p} n_i\ln n_i)$ under the constraints $\sum n_i =
N$ and $n_i \geq 0$.

It is straightforward to see that an optimal solution is obtained if
the computation time is the same for all processors (if a processor
ends its computation before another one, it could have been assigned
more work thus shortening the computation time of the busiest
processor).  The problem becomes to compute the data sizes
$n'_1,\cdots,n'_p$ such that:
\begin{equation}
n'_1 + n'_2 + \cdots + n'_p = N
\end{equation}
and such that
\begin{equation}\label{eq:equal}
(n'_1/k_1)\ln n'_1 = (n'_2/k_2)\ln n'_2 =\cdots = (n_p'/k_p)\ln n'_p
\end{equation}
We have shown that this new distribution converges to
the initial distribution when $N$ tends to infinity. We have also
proved in \cite{sbac04} that a constant time solution based on Taylor
developments leads to the following solution:
\begin{equation}
n_i = \frac{k_i}{K}N +\epsilon_i,\ \ \  (1\leq i\leq p)
\ \mbox{where}\ \epsilon_i=\frac{N}{\ln N}\left[\frac{k_i}{K^2}\sum_{j=1}^p k_j\ln\left(\frac{k_j}{k_i}\right)\right]
\end{equation}%
and where $K$ is simply the sum of the $k_i$. These equations give the
sizes that we must have to install initially on each processors to
guaranty that the processors will terminate at the same time. The time
cost of computing one $k_i$ is ${\cal O}(p)$ and is independent of $n$
which is an adequate property for the implementations since $p$ is
much lower and not of the same order than $n$.

One limitation of above the technique is that we assume that the cost
time of the code following the partitioning step should admit a Taylor
development. We introduce now a more general approach to solve the
problem of partitioning data in an heterogeneous context. It is the
central part of the work. We consider an analytic description of the
partitioning when the processors are uniformly related: processor $i$
has an intrinsic relative speed $k_i$.

\section{General exact analytic approach on uniformly related processors}\label{sec:analytic}

The problem we solved in past sections is to distribute batches of
size $N$ according to \eqref{eq:equal}. We will first replace the
execution time of the sorting function by a generic term $f(n)$ (which
would be $f(n)=n\ln n$ for a sorting function, but could also be
$f(n)=n^2$ for other sorting algorithms, or any function corresponding
to different algorithms).  We assume that $f$ is a strictly increasing
monotonous integer function. We can with this consider a more general
approach to task distribution in parallel algorithms. Since our
processors have an intrinsic relative speed $k_i$, the computation
time of a task of size $n_i$ will be $f(n_i)/k_i$. This (discrete)
function can be extended to a (real) function $\tilde{f}$ by
interpolation. We can try to solve this equation exactly through
analytical computation. We define the common execution time $T$
through the following equation:
\begin{equation}
T = \frac{\tilde{f}(n_1)}{k_1} = \frac{\tilde{f}(n_2)}{k_2} = \cdots =  \frac{\tilde{f}(n_p)}{k_p}
\end{equation}
and equation 
\begin{equation}\label{eq:all}
n_1 + n_2 + .... + n_p = N
\end{equation}
Let us recall that monotonous increasing functions can have an inverse
function. Therefore, for all $i$, we have $\tilde{f}(n_i)=T k_i$, and
thus:
\begin{equation}\label{eq:ni}
n_i=\tilde{f}^{-1}(T k_i)
\end{equation}
Therefore, we can rewrite \eqref{eq:all} as:
\begin{equation}\label{eq:solveT}
\sum_{i=1}^{p}\tilde{f}^{-1}(T k_i)=N
\end{equation}
If we take our initial problem, we have only one unknown term in this
equation which is $T$. The sum $\sum_{i=1}^{p}\tilde{f}^{-1}(T k_i)$
is a strictly increasing function of $T$. If we suppose $N$ large
enough, there is a unique solution for $T$. The condition of $N$ being
large enough is not a rough constraint. $\tilde{f}^{-1}(T)$ is the
number of data that can be treated in time $T$ by a processor speed
equals to $1$. If we consider that $\tilde{f}^{-1}(0)=0$ (which is
reasonable enough), we obtain that $\sum_{i=1}^{p}\tilde{f}^{-1}(T
k_i)=0$ for $T=0$.

Having $T$, it is easy to compute all the values of
$n_i=\tilde{f}^{-1}(T k_i)$. We shall show later on how this can be
used in several contexts. Note also that the computed values have to
be rounded to fit in the integer numbers. If the numbers are rounded
down, at most $p$ elements will be left unassigned to a processor. The
processors will therefore receive a batch of size
$n_i=\left\lfloor\tilde{f}^{-1}(T k_i)\right\rfloor+\tilde{\delta_i}$
to process. $\delta_i$ can be computed with the following (greedy)
algorithm:
\begin{enumerate}
\item Compute initial affectations
$\tilde{n}_i=\left\lfloor\tilde{f}^{-1}(T k_i)\right\rfloor$ and set
$\delta_i=0$;
\item For each unassigned item of the batch of size $N$ (at most $p$ 
elements) do:
\begin{enumerate}
\item Choose $i$ such that $(\tilde{n}_i+\delta_i+1)/k_i$ is the smallest;
\item Set $\delta_i=\delta_i+1$.
\end{enumerate}
\end{enumerate}

The running time of this algorithm is $O(p\log p)$ at most, so independant of the size of the data $N$.

\subsection{Multiplicative cost functions}

Let us consider now yet another cost function.  $f$ is a
multiplicative function if it verifies $f(x y)=f(x)f(y)$. If $f$ is
multiplicative and admits an inverse function $g$, its inverse is also
multiplicative:
\begin{equation*}
g(a b)=g(f(g(a))f(g(b)))=g(f(g(a)g(b)))=g(a)g(b)
\end{equation*}
If $\tilde{f}$ is such a function (e.g. $f(n)=n^k$), we can solve
equation~\eqref{eq:solveT} as follows:
\begin{equation}
N=\sum_{i=1}^{p}\tilde{f}^{-1}(T
k_i)=\sum_{i=1}^{p}\tilde{f}^{-1}(T)\tilde{f}^{-1}(k_i)=\tilde{f}^{-1}(T)\sum_{i=1}^{p}\tilde{f}^{-1}(k_i)
\end{equation}
We can then extract the value of $T$:
\begin{equation}
\tilde{f}^{-1}(T)=\frac{N}{\sum_{i=1}^{p}\tilde{f}^{-1}(k_i)}
\end{equation}
Combining it with~\eqref{eq:ni} we obtain:
\begin{equation}
n_i=\tilde{f}^{-1}(T k_i)=\tilde{f}^{-1}(T)\tilde{f}^{-1}(k_i)=\frac{\tilde{f}^{-1}(k_i)}{\sum_{i=1}^{p}\tilde{f}^{-1}(k_i)}N
\end{equation}

Hence the following result:
\begin{theorem}
If $f$ is a cost function with the multiplicative property $f(a
b)=f(a)f(b)$, then the size of the assigned sets is proportional to the
size of the global batch with a coefficient that depends on the relative
speed of the processor $k_i$:
$$n_i=\frac{\tilde{f}^{-1}(k_i)}{\sum_{i=1}^{p}\tilde{f}^{-1}(k_i)}N$$
\end{theorem}

This results is compatible with the usual method for linear
functions (split according to the relative speeds), and gives a nice
generalization of the formula.

\subsection{Sorting: the polylogarithmic function case}

Many algorithms have cost functions that are not multiplicative.  This
is the case for the cost $\Theta(n \log n)$ of the previous sequential
part of our sorting algorithm, and more generally for polylogarithmic
functions.  However, in this case equation \ref{eq:solveT} can be
solved numerically. Simple results show that polylogarithmic functions
do not yield a proportionality constant independent of $N$.

\subsubsection{Mathematical resolution for the case $n\ln n$}

In the case $f(n)=n\ln n$, the inverse function can be computed. It
makes use of the Lambert W function $W(x)$, defined as being the inverse
function of $x e^x$. The inverse of $f:n\mapsto n\ln n$ is therefore
$g:x\mapsto x/W(x)$.

The function $W(x)$ can be approached by well-known formulas, including
the ones given in~\cite{CJK97}. A development to the second order of the
formula yields $W(x)=\ln x-\ln\ln(x)+o(1)$, and also:
$$
\frac{x}{W(x)}=\frac{x}{\ln(x)}\frac{1}{1-(\ln\ln(x)/\ln(x))+o(1)}=
\frac{x}{\ln(x)}\left(1+{\frac{\ln\ln(x)}{\ln(x)}}+{\mathcal O}\left(\left({\frac{\ln\ln(x)}{\ln(x)}}\right)^2\right)\right)
$$
This approximation leads us to the following first-order approximation
that can be used to numerically compute in ${\mathcal O}(p)$ the value
of $T$:
\begin{theorem}\label{mytheo}
Initial values of $n_i$ can be asymptotically computed by
$$
\sum_{i=1}^{p}\frac{Tk_i+Tk_i\ln\ln(Tk_i)}{(\ln(Tk_i))^2}=N
\text{ and }
n_i=\frac{Tk_i+Tk_i\ln\ln(Tk_i)}{(\ln(Tk_i))^2}
$$
\end{theorem}

%

\subsection{Unknown cost functions}

Our previous method also claims an approach to unknown cost
functions. The general outline of the method is laid out, but needs
refinement according to the specific needs of the software platform.
When dealing with unknown cost functions, we assume no former
knowledge of the local sorting algorithm, just linear speed
adjustments (the collection of $k_i$). We assume however that the
algorithm has a cost function, i.e. a monotonous increasing function
of the size of the data $C$.%
\footnote{If some chunks are treated faster than smaller ones, their
complexity will be falsely exaggerated by our approach and lead to
discrepancies in the expected running time.} Several batch of data are
submitted to our software. Our method builds an incremental model of
the cost function. At first, data is given in chunks of size
proportionnal to each node's $k_i$. The computation time on node $i$
has a duration of $T_{n_i}$ and thus a basic complexity of
$C(n_i)=T_{n_i}k_i$. We can thus build a piecewise affine function (or more complex interpolated function, if heuristics require that) that represents the current knowledge of the system about the time
cost $n\mapsto C(n)$. Other values will be computed by
interpolation. The list of all \emph{known points} can be sorted, to
compute $f$ efficiently.

The following algorithm is executed for each task:
\begin{enumerate}
\item For each node $i$, precompute the mapping $(T,i)\mapsto n_i$ as previously, using interpolated values for $f$ if necessary (see below). Deduce a mapping $T\mapsto n$ by summing the mappings over all $i$.
\item Use a dichotomic search through $T\mapsto n$ mapping to find the ideal value of $T$ (and thus of all the $n_i$) and assign chunks of data to node $i$;
\item When chunk $i$ of size $n_i$ is being treated:
\begin{enumerate}
\item Record the cost $C=T_{n_i}k_i$ of the computation for size $n_i$.
\item If $n_i$ already had a non-interpolated value, choose a new value
$C'$ according to whatever strategy it fits for the precise platform
and desired effect (e.g. mean value weighted by the occurrences of the
various $C$ found for $n_i$, mean value weighted by the complexity of
the itemset, max value). Some strategies may require storing more
informations than just the mapping $n\mapsto C(n)$.
\item If $n_i$ was not a known point, set $C'=C$.
\item Ensure that the mapping as defined by $n\neq n_i\mapsto C(n)$
and the new value $n_i\mapsto C'$ is still monotonous increasing. If
not, raise or lower values of neighboring known points (this is simple
enough to do if the strategy is to represent the cost with a piecewise
function). Various heuristics can be applied, such as using the
weighted mean value of conflicting points for both points.
\end{enumerate}
\item At this point, the precomputation of the mappings will yield
consistent results for the dichotomic search. A new batch can begin.
\end{enumerate}

The initial extrapolation needs care. An idea of the infinite behavior
of the cost function toward infinity is a plus. In absence of any
idea, the assumption that the cost is linear can be a starting point
(a ``linear guess''). All ``linear guesses'' will yield chunks of data
of the same size (as in equation~\eqref{eq:equal}). Once at least one
point has been computed, the ``linear guess'' should use a ratio based
on the complexity for the largest chunk size ever treated (e.g. if size
$1,000$ yields a cost of $10,000$, the linear ratio should be at least
$10$).

\section{A dynamic programming technique for non-uniformly related processors}\label{sec:dynamic}

In the previous sections we have developed new constant time
solution to estimate the amount of data that each processor should
have in its local memory in order to ensure that the parallel sorts
end at the same time. The complexity of the method is the same than
the complexity of the method introduced in \cite{sbac04}. 

The class of functions that can be used according to the new method
introduced in the paper is large enough to be useful in practical cases. In
\cite{sbac04}, the class of functions captured by the method is the
class of functions that admit a Taylor development. It could be a
limitation of the use of the two methods.

Moreover, the approach of \cite{sbac04} considers that the processor
speeds are uniformly related, i.e. proportional to a given constant.
This is a restriction in the framework of heterogeneous computers
since the time to perform a computation on a given processor depends
not only on the clock frequency but also on various complex factors
(memory hierarchy, coprocessors for some operations).


In this section we provide a general method that provides an optimal
partitioning $n_i$ in the more general case. This method is based on
dynamic programming strategy similar to the one used in FFTW to find
the optimal split factor to compute the FFT of a vector \cite{FFTW05}.

Let us give some details of the dynamic approach. Let $f_i(m)$ be the
computational cost of a problem of size $m$ on machine $i$.  Note that
two distinct machines may implement different algorithms
(e.g. quicksort or radix sort) or even the same generic algorithm but
with specific threshold (e.g. Musser sort algorithm with processor
specific algorithm to switch from quicksort to merge sort and
insertion sort).  Also, in the sequel the $f_i$ are not assumed
proportional.

Given $N$, an optimal partitioning $(n_1, \ldots, n_p)$ with
$\sum_{i=1}^p n_i = N$ is defined as one that minimizes the parallel
computation time $T(N,p)$;
$$ 
 T(N,p) = \max \{{\vtop{\hbox{\hfill$f_i( n_i);\hfill$}\hbox{\raise 2pt\hbox{$\scriptstyle i=1,\ldots, p$}}}}\} =  
\min_{(x_1, \ldots, x_p)\in \mathbb{N}^p : \sum_{i=1}^p x_i  = N}
           \max \{{\vtop{\hbox{\hfill$f_i(x_i);\hfill$}\hbox{\raise 2pt\hbox{$\scriptstyle i=1,\ldots, p$}}}}\}
$$
%
%
A dynamic programming approach leads to the following inductive
characterization of the solution:$\forall (m,i) \mbox{~with~} 0 \leq m
\leq N \mbox{~and~} 1 \leq i \leq p: T(m,i) = \min_{n_i=0..m} \max (
f_i(n_i), C(m-n_i, i-1))$

Then, the computation of the optimal time $T(N,p)$ and of a related
partition $(n_i)_{i=1,\ldots, p}$ is obtained iteratively in ${\cal
O}(N^2.p)$ time and ${\cal O}(N.p)$ memory space.

The main advantage of the method is that it makes no assumption on the
functions $f_i$ that are non uniformly related in the general case.
Yet, the potential drawback is the computational overhead for
computing the $n_i$ which may be larger than the cost of the parallel
computation itself since $T(N,p) =o(N^2p)$.  However, it can be
noticed, as in \cite{FFTW05}, that this overhead can be amortized if
various input data are used with a same size $N$.  Moreover, some
values $T(m,p)$ for $m \leq K$ may be precomputed and stored.  Than in
this case, the overhead decreases to $O\left(p.\left(
\frac{N}{K}\right)^2\right)$.  Sampling few values for each $n_i$
enables to reduce the overhead as desired, at the price of a loss of
optimality.
 

\section{Experiments}\label{sec:experiments}

We have conducted experiments on the Grid-Explorer platform in order
to compare our approach for partitioning with partitioning based only
on the relative speeds. Grid-Explorer\footnote{See: {\tt
http://www.lri.fr/\~{}fci/GdX}} is a project devoted to build a large
scale experimental grid. The Grid-Explorer platform is connected also
to the nation wide project Grid5000\footnote{See: {\tt
http://www.grid5000.fr}} which is the largest Grid project in
France. We consider here only the Grid-Explorer platform which is
built with bi-Opteron processors (2Ghz, model 246), 80GB of IDE disks
(one per node). The interconnection network is made of Cisco switches
allowing a bandwidth of 1Gb/s full-duplex between any two
nodes. Currently, the Grid-Explorer platform has 216 computation
nodes (432 CPU) and 32 network nodes (used for network emulation -
not usefull in our case). So, the platform is an homogeneous
platform.

For emulating heterogeneous CPU, two techniques can be used. One can
use the CPUfreq driver available with Linux kernels (2.6 and above)
and if the processor supports it; the other one is CPU burning. In this
case, a thread with high priority is started on each node and consumes
Mhz while another process is started for the main program. In our
case, since we have bi-opteron processors we have chosen to run 2
processes per node and doing CPU burning, letting Linux to run them
one per CPU. Feedback and experience running the CPUfreq driver on a
bi-processor node, if it exists, is not frequent. This explain why we
use the CPU burning technique.

\begin{figure}[hbpt]
\begin{center}
  \includegraphics[width=80mm]{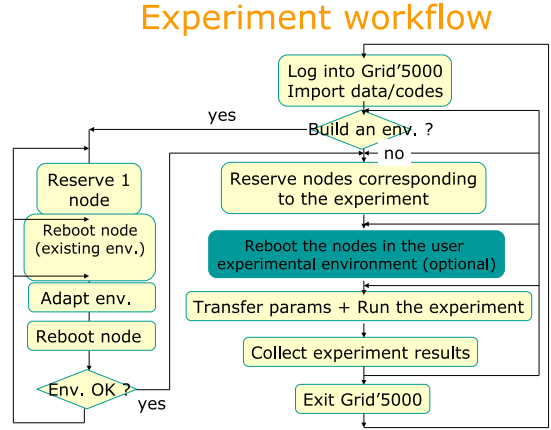}
\end{center} 
\label{FIGONE}
\caption{Methodology of experiments on the Grid-Explorer platform}
\end{figure}

Figure \ref{FIGONE} shows the methodology of running experiments on
the Grid-Explorer or Grid5000 platforms. Experimenters take care of
deploying codes and reserve nodes. After that, they configure an
environment (select specific packages and a Linux kernel, install
them) and reboot the nodes according to the environment. The
experiments take place only after installing this ``software stack''
and at a cost which is significant in term of time.
We have implemented the sorting algorithm depicted in subsection
\ref{mysection} and according to Theorem \ref{mytheo} for the
computation of the initial amount of data on each node for minimizing
the total execution time. Note that each node generates its local
portion on the local disk first, then we start to measure the time. It
includes the time for reading from disk, the time to select and to
exchange the pivots, the time for partitioning data according to the
pivots, the time for redistributing (in memory) the partitions, the
time for sorting and finally the time to write the result on the local
disks.

We sort records and each record is 100 bytes long. The first
10 bytes is a random key of 10 printable characters. We are compliant
with the requirements of Minute Sort\footnote{See: {\tt
http://research.microsoft.com/barc/SortBenchmark/}} as much as
possible in order to beat the record in a couple of weeks.

%



We proceed with 50 runs per experiment. We only consider here
experiments with a ratio of 1.5 between processor speeds. This is a
strong constraint: the more the ratio is high the more the difference
in execution time is important and in favor of our algorithm.  So we
have two classes of processor but the choice between the performance
($1$ or $1/1.5$) is made at random. We set half of the processors with
a performance of $1$ and the remainder with a performance of
$1/1.5$. We recall that the emulation technique is 'CPU burning'. 

Since we have observed that the communication time has a significant impact on
the total execution time, we have developed two strategies and among them, one
for using the second processor of the nodes. In the first implementation
communication take place in a single thread that is to say in the thread also
doing computation. In the second implementation we have created a thread for
sending and a thread for receiving the partitions. We guess that the operating
system allocates them on the 'second' processor of our bi-opteron cards
equiped with a single Gigabit card for communication.


The input size is 541623000 records (54GB) because it provides about
an execution time of one minute in the case of an homogeneous run
using the entire 2Ghz. Note that it corresponds approximatively to
47\% of the size of the 2005 Minute Sort record.



We run 3 experiments. Only experiments A.2 et A.3 use our technique to
partition the data whereas experiment A.1 corresponds to a
partitioning according to the relative speed only. In other words,
experiment A.1 corresponds to the case where the CPU burns X.Mhz
(where X is either 1Ghz or 1/1.5 GHz) but the performance vector is
set according to an homogeneous cluster, we mean without using our
method for re-balancing the work. Experiment A.2 also corresponds to
the case where communication are not done in separate threads (and
thus they are done on the same processor). Experiment A.3 corresponds
to the case where the CPU burns X Mhz (also with X is either 1Ghz or
1/1.5 GHz) and communication are done in separate threads (and thus
they are done on separate processors among the two available on a
node). We use the {\tt pthread} library and LAM-MPI 7.1.1 which is a
safe-thread implementation of MPI.  \begin{figure} \begin{center}
\begin{tabular}{ccc} \hline A1 experiment & A2 experiment & A3
experiment\\ \hline 125.4s & 112.7s & 69.4s\\ \hline \end{tabular}
\end{center} \label{mytable} \caption{Summary of experiments}
\end{figure} 
sorting 54GB on 96 nodes is depicted in Figure 2.  We observe that the
multithreaded code (A.3) for implementating the communication step is
more efficient than the code using a single thread (A.2).  This
observation confirms that the utilization of the second processor is
benefit for the execution time. Concerning the data partitioning
strategy introduced in the paper, we observe a benefit of about 10\%
in using it (A.2) comparing to A.1. Moreover, A.3 and A.2 use the same
partitioning step but they differ in the communication step. The
typical cost of the communication step is about 33\% of the execution
time for A.3 and about 60\% for A.2.

\section{Conclusion}\label{sec:conclusion}

In this paper we address the problem of data partitioning in
heterogeneous environments when relative speeds of processors are
related by constant integers. We have introduced the sorting problem
in order to exhibit inherent difficulties of the general problem.

We have proposed new ${\cal O}(p)$ solutions for a large class of time
complexity functions. We have also mentioned how dynamic programming can find
solutions in the case where cost functions are ``unrelated'' (we cannot depict
the cpu performance by the mean of integers) and we have reminded a recent and
promising result of Lastovetsky and Reddy related to a geometrical
interpretation of the solution. We have also described methods to deal with
unknown cost functions. Experiments based on heteroneous processors correlated
by a factor of 1.5 and on a cluster of 96 nodes (192 AMD Opteron 246) show
better performance with our technique compared to the case where processors
are supposed to be homogeneous. The performance of our algorithm is even
better if we consider higher factor for the heterogeneity notion,
demonstrating the validity of our approach.

In any case, communication costs are not yet taken into account. It is
an important challenge but the effort in modeling seems important. In
fact you cannot mix, for instance, information before the partitioning
with information after the partitioning in the same
equation. Moreover, communications are difficult to precisely modelize
in a complex grid archtitecture, where various network layers are
involved (Internet/ADSL, high speed networks,\ldots).  In this
context, a perspective is to adapt the static partitioning, such as
proposed in this paper, by a dynamic on-line redistribution of some
parts of the pre-allocated chunks in reaction to network overloads and
resources idleness (e.g. distributed work stealing).

\bibliographystyle{splncs}
\bibliography{these} 

\end{document}